\documentclass[12pt,a4paper]{article}
\usepackage{amsmath}
\usepackage{amsfonts}
\usepackage{amssymb}
\usepackage{makeidx}
\usepackage[applemac]{inputenc}

\begin{document}
\sloppy

\title{\textbf{The thermodynamic hamiltonian for open systems}}
\author{Umberto Lucia\\I.T.I.S. ``A. Volta'', Spalto Marengo 42, 15121 Alessandria, Italy}
\date{}
\maketitle

\begin{abstract}
The variational method is very important in mathematical and theoretical physics because it allows us to describe the natural systems by physical quantities independently from the frame of reference used. A global and statistical approach have been introduced starting from non-equilibrium thermodynamics, obtaining the principle of maximum entropy generation for the open systems. This principle is a consequence of the lagrangian approach to the open systems. Here it will be developed a general approach to obtain the thermodynamic hamiltonian for the dynamical study of the open systems. It follows that the irreversibility seems to be the fundamental phenomenon which drives the evolution of the states of the open systems.

\end{abstract}

\section{Introduction}
The variational method is very important in mathematical and theoretical physics because it allows us to describe the natural systems by physical quantities independently from the frame of reference used (Ozisik 1980). Moreover Lagrangian formulation can be used in a variety of physical phenomena and a structural analogy between different physical phenomena has been pointed out (Truesdell 1970). The most important result of the variational principle consists in obtaining both local and global theories (Truesdell 1970, Lucia 1995): global theory allows us to obtain information directly about the mean values of the physical quantities, while the local one about their distribution (Lucia 1995 - Lucia 2007). 

The notions of entropy and its production in equilibrium and non-equilibrium processes form the basis of modern thermodynamics and statistical physics (Bruers 2006, Dewar 2003, Maes \& Tasaki 2007, Martyushev \& Seleznev  2006, Wang 2007). Entropy has been proved to be a quantity describing the progress of non-equilibrium dissipative process. Great contribution has been done in this by Clausius, who in 1854-1862 introduced the notion of entropy in physics and by Prigogine who in 1947 proved the minimum entropy production principle (Martyushev \& Seleznev  2006). A Lagrangian approach to this subject allowed to obtained the mathematical consequences on the behaviour of the \textit{entropy generation}, elsewhere called \textit{entropy variation due to irreversibility} or \textit{irreversible entropy} $S_{irr}$ (Lucia 2008).

Theoretical and mathematical physics study idealized systems and one of the open problems is the understanding of how real systems are related to their idealization. The aim of this paper is to obtain a general rational thermodynamic approach to the analysis of irreversible systems and to develop the application of the entropy generation to the study of these systems by the dynamical approach. To do this we will introduce in Section 2 the thermodynamic open system (real system), in Section 3 the relation between the thermodynamics Lagrangian and the entropy generation, in Section 4 the thermodynamic hamiltonian and the hamiltonian approach to thermodynamics.

\section{The system considered}
In this section it will be defined the system considered. To do so, we must consider the definition of `system with perfect accessibility', which allows us to define both the thermodynamic system and the dynamical system.
\newtheorem{theorem}{Theorem}
\newtheorem{definition}{Definition}
Let us consider an open continuum or discrete $N$ particles system. Every $i-$th element of this system is located by a position vector $\mathbf{x}_{i}\in \mathbb{R}^{3}$, it has a velocity $\Dot{\mathbf{x}}_{i}\in \mathbb{R}^{3}$, a mass $m_{i}\in \mathbb{R}$ and a momentum $\mathbf{p}=m_{i}\mathbf{\Dot{x}}_{i}$, with $i\in [1,N]$ and $\mathbf{p}\in \mathbb{R}^{3}$ (Lucia 1995, Lucia 1998, Lucia 2001). The masses $m_{i}$ must satisfy the condition:
	\begin{equation}
		\sum_{i=1}^{N}m_{i}=m
	\end{equation} 
where $m$ is the total mass which must be a conserved quantity so that it follows:
	\begin{equation}
		\Dot{\rho}+\rho \nabla\cdot\Dot{\mathbf{x}}_{B}=0
	\end{equation} 
where $\rho =dm/dV$ is the total mass density, with $V$ total volume of the system and $\Dot{\mathbf{x}}_{B}\in \mathbb{R}^{3}$, defined as $\Dot{\mathbf{x}}_{B}=\sum_{i=1}^{N}\mathbf{p}_{i}/m$, velocity of the centre of mass. The  mass density must satisfy the following conservation law (Lucia 1995, Lucia 1998, Lucia 2001):
	\begin{equation}
		\Dot{\rho}_{i}+\rho_{i} \nabla\cdot\Dot{\mathbf{x}}_{i}=\rho \Xi
	\end{equation} 
where $\rho_{i}$ is the density of the $i-$th elementary volume $V_{i}$, with $\sum_{i=1}^{N}V_{i}=V$, and $\Xi$ is the source, generated by matter transfer, chemical reactions and thermodynamic transformations. This open system can be mathematical defined as follows (Lucia 2008).

\begin{definition} \emph{(Huang 1987)}
 - A dynamical state of $N$ particles can be specified by the $3N$ canonical coordinates $\left\{\mathbf{q}_{i}\in \mathbb{R}^{3},i\in \left[1,N\right]\right\}$ and their conjugate momenta $\left\{\mathbf{p}_{i}\in \mathbb{R}^{3},i\in \left[1,N\right]\right\}$. The $6N-$dimensional space spanned by $\left\{\left(\mathbf{p}_{i},\mathbf{q}_{i}\right),i\in \left[1,N\right]\right\}$ is called the phase space $\Omega$. A point $\mathbf{\sigma}_{i} =\left(\mathbf{p}_{i},\mathbf{q}_{i}\right)_{i\in \left[1,N\right]}$ in the phase space $\Omega :=\left\{\mathbf{\sigma}_{i}\in \mathbb{R}^{6N}:\mathbf{\sigma}_{i}=\left(\mathbf{p}_{i},\mathbf{q}_{i}\right), i\in \left[1,N\right]\right\}$ represents a state of the entire $N-$particle system.
\end{definition}

\begin{definition} \emph{(Lucia 2001)}
 - A system with perfect accessibility $\Omega_{PA}$ is a pair $\left(\Omega, \Pi\right)$, with $\Pi$ a set whose elements  $\pi$ are called process generators, together with two functions:
\begin{equation}
\pi \mapsto \mathcal{S}
\end{equation}
\begin{equation}
\left(\pi^{'},\pi{''}\right) \mapsto \pi^{''}\pi{'}
\end{equation}
where $\mathcal{S}$ is the state transformation induced by $\pi$, whose domain $\mathcal{D}\left(\pi\right)$ and range $\mathcal{R}\left(\pi\right)$ are non-empty subset of $\Omega$.
This assignment of transformation to process generators is required to satisfy the following conditions of accessibility:
\begin{enumerate}
	\item $\Pi\sigma:=\left\{\mathcal{S}\sigma :\pi\in\Pi,\sigma\in \mathcal{D}\left(\pi\right)\right\}=\Omega$ , $\forall \sigma\in\Omega$\emph{:} the set $\Pi\sigma$ is called the set of the states accessible from $\sigma$ and, consequently, it is the entire \emph{state space}, the phase spase $\Omega$;
	\item if $\mathcal{D}\left(\pi ''\right)\cap \mathcal{R}\left(\pi '\right)\neq 0\Rightarrow \mathcal{D}\left(\pi ''\pi '\right)=\mathcal{S}_{\pi '}^{-1}\left(\mathcal{D}\left(\pi ''\right)\right)$ and $\mathcal{S}_{\pi ''\pi '}\sigma =\mathcal{S}_{\pi ''}\mathcal{S}_{\pi '}\sigma$ $\forall\sigma\in \mathcal{D}\left(\pi ''\pi '\right)$
\end{enumerate}
\end{definition}

\begin{definition} \emph{(Lucia 2001)} 
- A process in $\Omega_{PA}$ is a pair $\left(\pi ,\sigma\right)$, with $\sigma$ a state and $\pi$ a process generator such that $\sigma$ is in $\mathcal{D}\left(\pi\right)$. The set of all processes of $\Omega_{PA}$ is denoted by:
\begin{equation}
\Pi \diamond\Omega =\left\{\left(\pi ,\sigma\right): \pi \in \Pi ,\sigma\in \mathcal{D}\left(\pi\right)\right\}
\end{equation}
If $\left(\pi ,\sigma\right)$ is a process, then $\sigma$ is called the initial state for $\left(\pi ,\sigma\right)$ and $\mathcal{S}\sigma$ is called the final state for $\left(\pi ,\sigma\right)$.
\end{definition}

\begin{definition} \emph{(Lucia 2001)} 
\label{thermdef} - A thermodynamic system is a system with perfect accessibility $\Omega_{PA}$ with two actions $W\left(\pi ,\sigma\right)\in\mathbb{R}$ and $H\left(\pi ,\sigma\right)\in\mathbb{R}$, called work done and heat gained by the system during the process $\left(\pi ,\sigma\right)$, respectively.
\end{definition}

\newtheorem{comment}{Comment}
\begin{comment} \emph{(Lucia 2008)} 
- The set of all these stationary states of a system $\Omega_{PA}$ is called non-equilibrium ensemble.
\end{comment}

\begin{definition} \emph{(Lucia 2001)}
- A thermodynamic path $\gamma$ is an oriented piecewise continuously differentiable curve in $\Omega_{PA}$.
\end{definition}

\begin{definition} \emph{(Lucia 2001)} 
- A cycle $\mathcal{C}$ is a path whose endpoints coincide.
\end{definition}

\begin{definition} \emph{(Billingsley 1979)}
 - A family $\mathcal{F}$ of subset of a perfect accessibility phase-space $\Omega_{PA}$ is said to be an \emph{algebra} if the following conditions are satisfied:
\begin{enumerate}
	\item $\Omega_{PA} \in \mathcal{F}$
	\item $\Omega_{PA}^{0} \in \mathcal{F} \Rightarrow \Omega_{PA}^{0^{c}}\in \mathcal{F}$
	\item $\left\{\Omega_{PA_{i}}\right\}_{i\in \left[1,k\right]}\subset \mathcal{F} \Rightarrow \bigcup_{i=1}^{k} \Omega_{PA_{i}} \in \mathcal{F}$
\end{enumerate}
\end{definition}

\begin{comment} \emph{(Lucia 2008)} 
 - Moreover, it follows that:
\begin{enumerate}
	\item $\emptyset \in \mathcal{F}$
	\item the algebra $\mathcal{F}$ is closed under countable intersections and subtraction of sets
	\item if $k\equiv \left\{\infty \right\}$ then $\mathcal{F}$ is said a $\sigma-$algebra
\end{enumerate}
\end{comment}

\begin{definition} \emph{(Billingsley 1979)} 
 - A function $\mu :\mathcal{F}\rightarrow \left[0,\infty\right)$ is a measure if it is additive. It means that for any countable subfamily $\left\{\Omega_{i}, i\in\left[1,n\right]\right\} \subseteq\mathcal{F}$, consisting of mutually disjoint sets, it follows:
\begin{equation}
 \mu\left(\bigcup_{i=1}^{n}\Omega_{i}\right)=\sum_{i=1}^{n}\mu\left(\Omega_{i}\right)
\end{equation}
It follows, too:
\begin{enumerate}
	\item $\mu\left(\emptyset\right)=0$
	\item if $\Omega_{\alpha}, \Omega_{\beta} \in\mathcal{F}$ and $\Omega_{\alpha}\subset\Omega_{\beta}\Rightarrow\mu\left(\Omega_{\alpha}\right)\leq\mu\left(\Omega_{\beta}\right)$
	\item if $\Omega_{1}\subset\Omega_{2}\subset\dots\subset\Omega_{n}$ and $\left\{\Omega_{i},i\in\left[1,n\right]\right\}\in\mathcal{F}\Rightarrow\mu\left(\bigcup_{i=1}^{n}\Omega_{i}\right)=\sup_{i}\mu\left(\Omega_{i}\right)$.
\end{enumerate}
Moreover, if $\mathcal{F}$ is a $\sigma-$algebra and $n\equiv\left\{\infty\right\}$ then the measure is said $\sigma-$additive.
\end{definition}

\begin{definition} \emph{(Gallavotti 2003)}
- A smooth map $\mathcal{S}$ of a compact manifold $\mathcal{M}$ is a map with the property that around each point $\mathbf{\sigma}$ it can be established a system of coordinates based on smooth surfaces $W^{s}_{\mathbf{\sigma}}$ and $W^{u}_{\mathbf{\sigma}}$, with $s$=stable and $u$=unstable, of complementary positive dimension which is:
\begin{enumerate}
	\item \emph{covariant:} $\partial\mathcal{S} W^{i}_{\mathbf{\sigma}}=W^{i}_{\mathcal{S}\mathbf{\sigma}}, i=u,s$. This means that the tangent planes $\partial\mathcal{S} W^{i}_{\mathbf{\sigma}}, i=u,s$ to the coordinates surface at $\mathbf{\sigma}$ are mapped over the corresponding planes at $\mathcal{S}\mathbf{\sigma}$;
	\item \emph{continuous:} $\partial\mathcal{S} W^{i}_{\mathbf{\sigma}}$, with $i=u,s$, depends continuously on $\mathbf{\sigma}$;
	\item \emph{transitivity:} there is a point in a subsistem of $\Omega_{PA}$ of zero Liouville probability, called attractor, with a dense orbit.
\end{enumerate}
\end{definition}

\begin{comment} \emph{(Lucia 2008)} 
A great number of systems satisfies also the \emph{hyperbolic condition:} the lenght of a tangent vector $\mathbf{v}$ is amplified by a factor $C\lambda ^{k}$ for $k>0$ under the action of $\mathcal{S}^{-k}$ if $\mathbf{\sigma}\in W^{s}_{k}$ with $C>0$ and $\lambda <1$. This means that if an observer moves with the point $\mathbf{\sigma}$ it sees the nearby points moving around it as if it was a hyperbolic fixed point. But, in a general approach this condition is not necessary to be introduced.
\end{comment}

\newtheorem{hypothesis}{Hypothesis}
\begin{hypothesis} \emph{(Lucia 2008)} 
- There exists a statistics $\mu _{PA}$ describing the asymptotic behaviour of almost all initial data in perfect accessibility phase space $\Omega_{PA}$ such that, except for a volume zero set of initial data $\mathbf{\sigma}$, it will be:
\begin{equation}
\lim_{T\longrightarrow\infty}\frac{1}{T}\sum^{T-1}_{j=1}\varphi\left(\mathcal{S}^{j}\mathbf{\sigma}\right)=\int_{\Omega}\mu_{PA}\left(d\mathbf{\sigma}\right)\varphi\left(\mathbf{\sigma}\right)
\end{equation}
for all continuous functions $\varphi$ on $\Omega_{PA}$ and for every transformation $\mathbf{\sigma}\mapsto \mathcal{S}_{t}\left(\mathbf{\sigma}\right)$. For hyperbolic systems the distribution $\mu_{PA}$ is the Sinai-Ruelle-Bowen distribution, SRB-distribution or SRB-statistics \emph{(Gallavotti 1995)}.
\end{hypothesis}

\begin{comment}  \emph{(Lucia 2008)} 
- The notation $\mu_{PA}\left(d\mathbf{\sigma}\right)$ expresses the possible fractal nature of the support of the distribution $\mu$, and implies that the probability of finding the dynamical system in the infinitesimal volume $d\mathbf{\sigma}$ around $\mathbf{\sigma}$ may not be proportional to $d\mathbf{\sigma}$. Consequently, it may not be written as $\mu_{PA}\left(\mathbf{\sigma}\right) d\mathbf{\sigma}$, but it needs to be conventionally expressed as $\mu_{PA}\left(d\mathbf{\sigma}\right)$. The fractal nature of the phase space is an issue yet under debate \emph{(García-Morales \& Pellicer 2006)}, but there are a lot of evidence on it in the low dimensional systems \emph{(}see ref. \emph{García-Morales \& Pellicer 2006} and \emph{Hoover 1998}\emph{)}. Here we want to consider also this possibility.
\end{comment}

\begin{definition} \emph{(Lucia 2008)} 
 - The triple $\left(\Omega_{PA},\mathcal{F},\mu_{PA}\right)$ is a \emph{measure space}, the \emph{Kolmogorov probability space} $\Gamma$.
\end{definition}

\begin{definition} \emph{(Lucia 2008)} 
 - A dynamical law $\tau$ is a group of meausure-preserving automorphisms $\mathcal{S}:\Omega_{PA}\rightarrow\Omega_{PA}$ of the probability space $\Gamma$.
\end{definition}

\begin{definition} \emph{(Berkovitz \textit{et al.} 2006)}
 - A dynamical system $\Gamma_{d}=\left(\Omega_{PA},\mathcal{F},\mu_{PA},\tau\right)$ consists of a dynamical law $\tau$ on the probability space $\Gamma$.
\end{definition}

\section{Thermodynamic Lagrangian and entropy generation}
Theoretical and mathematical physics study idealized systems and one of the open problems is the understanding of how real systems are related to their idealization. In the analysis of irreversibility it has been introduced the concept of thermostats, which are systems of particles moving outside the system and interacting with the particles of the system through interactions across the walls of the system itself. In the statistical and dynamical approach to thermodynamics, the fundamental quantity considered by Gallavotti is the entropy production $\sigma_{entr}$, defined as follows (Gallavotti 2003):
\begin{equation}	\sigma_{entr}=k_{B}\sigma_{+}^{entr}=-k_{B}\int_{\Omega}\mu_{PA}\left(d\mathbf{\sigma}\right)\nabla\cdot \mathbf{E}\left(\mathbf{\sigma}\right)
	\label{entropy production}
\end{equation}
with always $\sigma_{+}^{entr}\geq 0$ and $\sigma_{+}^{entr}=0$ only for the equilibrium state and $\mathbf{E}\left(\mathbf{\sigma}\right):=\mathbf{f}_{int}\left(\mathbf{\sigma}\right)+\mathbf{f}_{nc}\left(\mathbf{\sigma}\right)+\mathbf{f}_{term}\left(\mathbf{\sigma}\right)$, with $\mathbf{f}_{int}\left(\mathbf{\sigma}\right)$ internal consevation force, $\mathbf{f}_{nc}\left(\mathbf{\sigma}\right)$ external active non conservative force and $\mathbf{f}_{term}\left(\mathbf{\sigma}\right)$ thermostatic expression.

\begin{comment} \emph{(Lucia 1995)}
- Following Truesdell \emph{(Truesdell 1970)}, for each continuum thermodynamic system (isolated or closed or open), in which it is possible to identify a thermodynamics subsystem with elementary mass $dm$ and elementary volume $dV=dm/\rho$, with $\rho$ mass density \emph{(Truesdell 1970, Lucia 1995, Lucia 1997, Lucia 2001, Lucia 2007)} the thermodynamic description can be developed by referring to the generalized coordinates $\left\{\xi_{i},\dot{\xi}_{i},t\right\}_{i\in[1,N]}$, with $\xi_{i} =\alpha_{i}-\alpha^{\left(0\right)}_{i}$, $\alpha_{i}$ the extensive thermodynamic quantities and $\alpha^{\left(0\right)}_{i}$ their values at the stable states.
\end{comment}

In thermodynamic engineering the energy lost for irreversible processes is evaluated by the first and second law of thermodynamics for the open systems (Lucia 1998) . So the following definition can be introduced:
\begin{definition} \emph{(Lucia 2008)} 
 - The entropy generation $\Delta S_{irr}$ is defined as:
\begin{equation}
\frac{W_{lost}}{T_{ref}}=\Delta S_{irr}:=\frac{Q_{r}}{T_{a}}\left( 1- \frac{T_{a}}{T_{r}} \right)+\frac{\Delta H}{T_{a}}-\Delta S+\frac{\Delta E_{k}+\Delta E_{g}-W}{T_{a}}
\end{equation}
where $W_{lost}$ is the work lost for irreversibility, $T_{ref}$ the temperature of the lower reservoir, $Q_{r}$ is the heat source, $T_{r}$ its temperature, $T_{a}$ is the ambient temperature, $H$ is the entalpy, $S$ is the entropy, $E_{k}$ is the kinetic energy, $E_{g}$ is the gravitational one and $W$ is the work.
\end{definition}

\begin{comment} \emph{(Lucia 2008)}
- It has been proved that $\Delta S_{irr}=\Dot{m}\sigma_{entr}$, with $\Dot{m}$ mass flow.
\end{comment}

\begin{theorem} 
\emph{(Lucia 1995, Lucia 1997, Lucia 2001, Lucia 2007, Lucia 2008)}
- The termodynamic Lagrangian can be obtained as:
\begin{equation}
	\mathcal{L}=-T_{ref}\Delta S_{irr}
		\label{Lagrangian}
\end{equation}
\end{theorem}
\textit{Proof}.
For every subsystem a termodynamic Lagrangian per unit time $t$, temperature $T$ and volume $V$, $\rho_{\mathcal{L}}$ is defined as (Lucia 1995, Lucia 1997, Lucia 2001, Lucia 2007):
\begin{equation}
	\rho_{\mathcal{L}}=\frac{d^{3}\mathcal{L}}{dV dT dt}=\frac{d\dot{S}}{dV}-\psi
		\label{Lagrangian density}
\end{equation}
where (Lucia 1995):
\begin{equation}
\frac{d\dot{S}}{dV}=\sum_{ij}L_{ij}\xi_{i}\xi_{j}+\frac{1}{2}\sum_{ijk}L_{ijk}\xi_{i}\xi_{j}\xi_{k}
		\label{entropy density}
\end{equation}
is the entropy per unit time and volume, and $\psi$ is the non-linear dissipative potential density, defined as (Lucia 1995):
\begin{equation}
\psi=\frac{1}{2}\sum_{ij}L_{ij}\xi_{i}\xi_{j}+\frac{1}{6}\sum_{ijk}L_{ijk}\xi_{i}\xi_{j}\xi_{k}
		\label{dissipative potential density}
\end{equation}
with $L_{ij}$ Onsager coefficients, defined as (Gallavotti 2006):
\begin{equation} 
L_{ij}=\frac{1}{2}\int_{-\infty}^{\infty}\mu_{PA}\left[\mathbf{\sigma}_{i}\left(\mathcal{S}_{t}\mathbf{\sigma}\right)\mathbf{\sigma}_{j}\left(\mathbf{\sigma}\right)\right]_{\mathbf{E}=\mathbf{0}}dt
\end{equation}
where $\mathbf{\sigma}_{\alpha}$, $\alpha =i,j$, are the $\alpha$-state and $\mathbf{E}=\left\{E_{k}\right\}_{k\in\left[1,r\right]}$ are the $r$ external forces.
Consequently, $\rho_{\mathcal{L}}$ becomes:
\begin{equation}
\rho_{\mathcal{L}}=\frac{1}{2}\sum_{ij}L_{ij}\xi_{i}\xi_{j}+\frac{1}{3}\sum_{ijk}L_{ijk}\xi_{i}\xi_{j}\xi_{k}
		\label{Lagrangian density expressio}
\end{equation}
which is the Legendre transformation to the relation (\ref{dissipative potential density}), too (Lucia 1995). Moreover, the thermodynamic Lagrangian $\mathcal{L}$ is defined as:
\begin{equation}
\mathcal{L}:=\int_{t}dt\int_{T}dT\int_{V}dV\rho_{\mathcal{L}}
\end{equation}
and considering that (Lucia 1995):
\begin{equation}
	\rho_{\mathcal{L}}=\rho_{S}-\rho_{\pi}-\psi
\end{equation}
where $\rho_{S}$ is the entropy per unit time and mass and $\rho_{\pi}$ is the power per unit mass and temperature. Now, following Lavenda (Lucia 2001), $\rho_{S}-\rho_{\pi}=2\psi$, so that (Lucia 2001, Lucia 2007):
\begin{equation}
	\rho_{\mathcal{L}}=\psi
\end{equation}
and
\begin{equation}
	\int_{t}dt\int_{T}dT\int_{V}dV\rho_{\mathcal{L}}=\int_{t}dt\int_{T}dT\int_{V}dV\psi
\end{equation}
so it follows:
\begin{equation}
\mathcal{L}=\int_{t}dt\int_{T}dT\int_{V}dV\psi
\end{equation}
but, remembering that:
\begin{equation}
W_{lost}:=\int_{t}dt\int_{T}dT\int_{V}dV\psi
\end{equation}
we can obtain:
\begin{equation}
\mathcal{L}=W_{lost}
\end{equation}
Considering the Gouy-Stodola theorem (Lucia 1995), which states that:
\begin{equation}
W_{lost}=-T_{ref}\Delta S_{irr}
\end{equation}
the (\ref{Lagrangian}) becomes:
\begin{equation}
	\mathcal{L}=\int_{t}dt\int_{T}dT\int_{V}dV \psi=-T_{ref}\Delta S_{irr}
		\label{Lagrangia}
\end{equation}
with $T_{ref}$ the temperature of the lower reservoir and $S_{irr}$ the entropy generation  (Lucia 2001, Lucia 2007).

\begin{definition} \emph{(Lucia 2008)}
The statistical expression, for the irreversible-entropy variation, results:
\begin{equation}
\Delta S_{irr}=-\frac{k_{B}}{\Dot{m}}\int_{\Omega}\mu_{PA}\left(d\mathbf{\sigma}\right)\nabla\cdot \mathbf{E}\left(\mathbf{\sigma}\right)
\label{irrentrdef}
\end{equation}
\end{definition}

\begin{theorem} \emph{(Lucia 2007)}
- \label{maxentrth} \textbf{The principle of maximum entropy generation} \emph{(Lucia 2007):}
 The condition of stability for the open system' stationary states is that its entropy generation $\Delta S_{irr}$ reaches its maximum:
\end{theorem}
\begin{equation}
\delta\left(\Delta S_{irr}\right)\geq 0
\label{maxent}
\end{equation}
\textit{Proof}. The thermodynamic action is defined as (Truesdell 1970):
\begin{equation}
\mathcal{A}:=\int_{t}dt \mathcal{L}
\end{equation}
From the principle of the least action:
\begin{equation}
\delta\mathcal{A}\leq 0
\end{equation}
remebering the relation (\ref{Lagrangian}) it follows that:
\begin{equation}
	-\delta\left(\int dt T_{ref}\Delta S_{irr}\right)\leq 0 
\end{equation}
which becomes:
\begin{equation}
\delta\left(T_{ref}\Delta S_{irr}\right)\geq 0
\end{equation}
and if $T_{ref}$ is constant
\begin{equation}
\delta\left(\Delta S_{irr}\right)\geq 0
\end{equation}

\section{Thermodynamic Hamiltonian and entropy generation}
\begin{theorem}
- \textbf{The thermodynamic hamiltonian}\emph{:}
 The thermodynamic Hamiltonian can be obtain as $T_{ref} \Delta S_{irr}$.
\end{theorem}
\textit{Proof}.
The thermodynamic Hamiltonian density $\rho_{\mathcal{H}}$ can be defined following the definition of the thermodynamic lagrangian density $\rho_{\mathcal{L}}$:
\begin{equation}
\rho_{\mathcal{H}}=\sum_i \zeta_i\xi_i - \rho_{\mathcal{L}}
\end{equation}
with
\begin{equation}
\zeta_i=\frac{\partial\rho_{\mathcal{L}}}{\partial\Dot{\xi}_i}
\end{equation}
coniugate generalised momentum to the generalised coordinates $\xi_i$. The thermodynamic lagrangian density $\rho_{\mathcal{L}}$ is defined by the relation (\ref{Lagrangian density}) and it is a function only of the generalised coordinates, so that it follows that:
\begin{equation}
\frac{\partial\rho_{\mathcal{L}}}{\partial\Dot{\xi}_i}=0
\label{momentum}
\end{equation}
Consequently, the thermodynamic Hamiltonian density results:
\begin{equation}
\rho_{\mathcal{H}}= - \rho_{\mathcal{L}}
\end{equation}
and the thermodynamic Hamiltonian can be obtained as:
\begin{equation}
\begin{split}
\mathcal{H}=\int_{t}dt\int_{T}dT\int_{V}dV \rho_{\mathcal{H}}=-\int_{t}dt\int_{T}dT\int_{V}dV \rho_{\mathcal{L}}=\\=-\mathcal{L}=T_{ref} \Delta S_{irr}
\end{split}
\end{equation}

\begin{comment}
From the definition of action it follows that the thermodynamic action can be written as:
\begin{equation}
\mathcal{A}=\int_{t}dt \mathcal{L}=-\int_{t}dt \mathcal{H}
\end{equation}
\end{comment}

\begin{comment}
The thermodynamic hamiltonian for open systems is related only to the entropy generation. Consequently, this quantity seems to be the basis of the analysis of these systems.
\end{comment}

\begin{comment}
Moreover, the irreversibility seems to be the fundamental phenomenon which drives the evolution of the states of the open systems.
\end{comment}

\begin{comment}
From the relation \emph{(\ref{momentum})} it follows that:
\begin{equation}
\zeta_i=0\Leftrightarrow\Dot{\zeta_i}=0\Leftrightarrow\frac{\partial\mathcal{H}}{\partial\xi_i}=0\Leftrightarrow\frac{\partial\rho_{\mathcal{H}}}{\partial\xi_i}=0
\end{equation}
and as a consequence of the relation \emph{(16)} it is possible to obtain:
\begin{equation}
\psi=\rho_{\mathcal{L}}=-\rho_{\mathcal{H}}=\frac{1}{6}\sum_{ijk}L_{ijk}\xi_i\xi_j\xi_k
\end{equation}
proving the completely non-linear behaviour of irreversibility and dissipation.
\end{comment}

\begin{comment}
From the relation \emph{(\ref{momentum})} it follows that:
\begin{equation}
\Dot{\xi}_i=\frac{\partial\rho_{\mathcal{H}}}{\partial\zeta_i}=0
\end{equation}
so:
\begin{equation}
\xi_i=\text{constant}
\end{equation}
proving that the dissipation not varies the velocity of the points inside the phase space, but it varies the path, in agreement with the hypothesis of Jaynes \emph{(Dewar, R. 2003)}.
\end{comment}

\section{Conclusions}
The aim of this paper was to obtain a general rational thermodynamic approach to the analysis of irreversible systems and to develop the application of the entropy generation to the study of these systems by the dynamical approach. To do this we will introduce in Section 2 the thermodynamic open system (real system), in Section 3 the relation between the thermodynamics Lagrangian and the entropy generation, in Section 4 the thermodynamic hamiltonian and the hamiltonian approach to thermodynamics.

The variational method is very important in mathematical and theoretical physics because it allows us to describe the natural systems by physical quantities independently from the frame of reference used (Ozisik 1980). Moreover Lagrangian formulation can be used in a variety of physical phenomena and a structural analogy between different physical phenomena has been pointed out (Truesdell 1970). The most important result of the variational principle consists in obtaining both local and global theories (Truesdell 1970, Lucia 1995): global theory allows us to obtain information directly about the mean values of the physical quantities, while the local one about their distribution (Lucia 1995 - Lucia 2007). 

In phenomena out of equilibrium irreversibility manifests itself because the fluctuations of the physical quantities, which bring the system apparently out of stationarity, occur symmetrically about their average values (Gallavotti 2006). The notions of entropy and its production in equilibrium and non-equilibrium processes form the basis of modern thermodynamics and statistical physics (Bruers 2006, Dewar 2003, Maes \& Tasaki 2007, Martyushev \& Seleznev  2006, Wang 2007). 

Entropy has been proved to be a quantity describing the progress of non-equilibrium dissipative process. Great contribution has been done in this by Clausius, who in 1854-1862 introduced the notion of entropy in physics and by Prigogine who in 1947 proved the minimum entropy production principle (Martyushev \& Seleznev  2006). A Lagrangian approach to this subject allowed to obtained the mathematical consequences on the behaviour of the \textit{entropy generation}, elsewhere called \textit{entropy variation due to irreversibility} or \textit{irreversible entropy} $S_{irr}$ (Lucia 2008). The $\varepsilon$-stady state definition allowed us to obtain that for certain fluctuations the probability of occurrence follows a universal law and the frequency of occurrence is controlled by a quantity that has been related to the entropy generation (Lucia 2008). Moreover, this last quantity has a purely mechanical interpretation which is related to the the ergodic hypothesis which proposed that an isolated system evolves in time visiting all possible microscopic states. Moreover, considering that the open system is a system with perfect accessibility represented as a probability space in which is defined a $PA$-measure and a statistical approach has been developed (Lucia 2008). To link this statistical approach to the dynamical one it is needed to obtain the thermodynamic Hamiltonian for the open systems.

It follows that the thermodynamic hamiltonian for open systems is related only to the entropy generation. Consequently, this quantity seems to be the basis of the analysis of these systems. Moreover, the irreversibility seems to be the fundamental phenomenon which drives the evolution of the states of the open systems, and that irreversibility and dissipation have a completely non-linear behaviour inside the phase space where the points of the open system' states move at constant velocity, but on particular path, not equal to the reversible systems, as proving the hypothesis introduced by Jaynes \emph{(Dewar, R. 2003)}


\end{document}